\documentclass[preprint,5p,times,twocolumn,longtitle,numbers,round,sort&compress]{elsarticle}
\usepackage[T1]{fontenc}
\usepackage[english]{babel}

\usepackage{amsmath}
\usepackage{graphicx}
\usepackage[colorlinks=true, allcolors=blue]{hyperref}
\usepackage[normalem]{ulem}
\usepackage{comment}

\newcommand{\sqrtsNN}{\ensuremath{\sqrt{s_{_\mathrm {NN}}}}}
\newcommand{\IAA}{\ensuremath{I_{\mathrm{AA}}}}
\newcommand{\ICP}{\ensuremath{I_{\mathrm{CP}}}}
\newcommand{\pT}{\ensuremath{p_\mathrm{T}}}

\newcommand{\pTjetch}{\ensuremath{p_{\mathrm{T,jet}}^{\mathrm{ch}}}}

\newcommand{\pTtrig}{\ensuremath{p_\mathrm{T,trig}}}

\newcommand{\Rbrtwofive}{\ensuremath{\mathfrak{R}^{0.2/0.5}}}
\newcommand{\Rbrthreefive}{\ensuremath{\mathfrak{R}^{0.3/0.5}}}
\newcommand{\Rbrfourfive}{\ensuremath{\mathfrak{R}^{0.4/0.5}}}

\newcommand{\zvtxTPC}{\ensuremath{V_\mathrm{z,TPC}}}
\newcommand{\zvtxVPD}{\ensuremath{V_\mathrm{z,VPD}}}

\newcommand{\Qsqr}{\ensuremath{Q^2}}

\newcommand{\AuAu}{Au+Au}
\newcommand{\OO}{O+O}
\newcommand{\RuRu}{Ru+Ru}
\newcommand{\ZrZr}{Zr+Zr}
\newcommand{\PbPb}{Pb+Pb}

\newcommand{\aaa}{A+A}

\newcommand{\pA}{\ensuremath{p\mathrm{+A}}}
\newcommand{\pPb}{\ensuremath{p\mathrm{+Pb}}}
\newcommand{\pp}{\ensuremath{p}+\ensuremath{p}}

\newcommand{\kT}{\ensuremath{k_\mathrm{T}}}
\newcommand{\antikT}{anti-\ensuremath{k_\mathrm{T}}}

\newcommand{\pizero}{\ensuremath{\pi^0}}

\newcommand{\Ajet}{\ensuremath{A_\mathrm{jet}}}
\newcommand{\pTraw}{\ensuremath{p_{\mathrm{T,jet}}^{\mathrm{raw,ch}}}}
\newcommand{\pTreco}{\ensuremath{p_{\mathrm{T,jet}}^{\mathrm{reco,ch}}}}
\newcommand{\dphi}{\ensuremath{\Delta\varphi}}
\newcommand{\Nch}{\ensuremath{N_{\mathrm{ch}}}}
\newcommand{\fME}{\ensuremath{f^{\rm ME}}}

\newcommand{\Rdet}{\ensuremath{R_{\rm det}}}
\newcommand{\Rbkg}{\ensuremath{R_{\rm bkg}}}
\newcommand{\Rtot}{\ensuremath{R_{\rm tot}}}
\newcommand{\pTpart}{\ensuremath{p_{\mathrm{T,jet}}^{\mathrm{part,ch}}}}
\newcommand{\pTdet}{\ensuremath{p_{\mathrm{T,jet}}^{\mathrm{det,ch}}}}

\newcommand{\YhJet}{\ensuremath{Y^\mathrm{h+jet}}}
\newcommand{\YtildehJet}{\ensuremath{\tilde{Y}^\mathrm{h+jet}}} 

\newcommand{\etajetch}{\ensuremath{\eta_{\mathrm{jet}}^\mathrm{ch}}}
\newcommand{\Ntrig}{\ensuremath{N_\mathrm{trig}}}
\newcommand{\Njetch}{\ensuremath{N_{\mathrm{jet}}^\mathrm{ch}}}

\journal{Physics Letters B}

\begin{document}


\begin{frontmatter}

\title{Measurements of jet quenching with semi-inclusive hadron-jet correlations in Ru+Ru and Zr+Zr collisions at $\sqrt{s_\mathrm{NN}}=200$ GeV}

\author{
	B.~E.~Aboona$^{59}$,
	J.~Adam$^{18}$,
	L.~Adamczyk$^{4}$,
	I.~Aggarwal$^{46}$,
	M.~M.~Aggarwal$^{46}$,
	Z.~Ahammed$^{68}$,
	A.~K.~Alshammri$^{34}$,
	E.~C.~Aschenauer$^{8}$,
	S.~Aslam$^{23}$,
	J.~Atchison$^{3}$
% ...
}

\address{\rm{(STAR Collaboration)}}

% ...
\address{$^{3}$Abilene Christian University, Abilene, Texas   79699}
\address{$^{4}$AGH University of Krakow, FPACS, Cracow 30-059, Poland}
% ...
\address{$^{8}$Brookhaven National Laboratory, Upton, New York 11973}
% ...
\address{$^{18}$Czech Technical University in Prague, FNSPE, Prague 115 19, Czech Republic}
% ...
\address{$^{23}$Fudan University, Shanghai, 200433 }
% ...
\address{$^{34}$Kent State University, Kent, Ohio 44242}
% ...
\address{$^{46}$Panjab University, Chandigarh 160014, India}
% ...
\address{$^{59}$Texas A\&M University, College Station, Texas 77843}
% ...
\address{$^{68}$Variable Energy Cyclotron Centre, Kolkata 700064, India}
% ...

\begin{abstract}
The STAR experiment at RHIC reports measurements of the semi-inclusive yield of charged-particle jets recoiling from high transverse momentum charged-hadron triggers in centrality-selected \RuRu\ and \ZrZr\ collisions at the nucleon–nucleon center-of-mass energy of 200 GeV. The effects of jet quenching, arising from the interaction of jets with the quark--gluon plasma, are quantified by comparing trigger-normalized recoil yields in central and peripheral collisions. Suppression of the recoil yield in central collisions is observed, indicating medium-induced partonic energy loss due to quenching. The ratio of recoil jet yields for small and large resolution parameter is found to be suppressed in central relative to peripheral collisions, characteristic of medium-induced intra-jet broadening. The results are compared to similar measurements in smaller and larger collision systems, providing unique insight into the system-size dependence of jet quenching.
\end{abstract}

\end{frontmatter}

\section{Introduction}

The quark--gluon plasma (QGP) is a state of strongly-interacting matter at very high temperature, in which quarks and gluons (partons) are deconfined~\cite{Busza:2018rrf,Elfner:2022iae,Harris:2023tti}. A QGP filled the early universe a few microseconds after the Big Bang~\cite{Schwarz:2003du}, and QGP is generated and studied today in high-energy nuclear collisions at the Relativistic Heavy Ion Collider (RHIC) and the Large Hadron Collider (LHC)~\cite{Harris:2023tti}. Comparison of collider data with theoretical model calculations shows that the QGP formed in such nuclear collisions is a near-perfect fluid, flowing with very low specific shear viscosity~\cite{Harris:2023tti,Heinz:2013th,2025-1834}.

In high-energy hadronic collisions, hard scatterings (large squared momentum transfer $Q^2$) of partons in the projectiles generate highly-virtual quarks and gluons, which manifest as collimated sprays of experimentally observable  hadrons (``jets'')~\cite{Sterman:1977wj,Salam:2010nqg}. In high-energy nuclear collisions, jet formation occurs concurrently with the generation and evolution of the QGP. Jets interact with the QGP, producing observable modifications in jet production rates, correlations, and internal structure (``jet quenching'')~\cite{Majumder:2010qh,Wang:2025lct}. Comparison of jet quenching measurements and theoretical model calculations provides unique and incisive probes of jet transport in the QGP, and the response of the QGP to excitation~\cite{Cunqueiro:2021wls,Apolinario:2022vzg,JET:2013cls,JETSCAPE:2024cqe,Jing:2025bwi}.

In the theoretical description of jet quenching, jet--QGP interactions are governed by long-range coherence effects that are the Quantum Chromodynamics (QCD) analogue to Landau--Pomeranchuk--Migdal (LPM) coherence effects on scattering in Quantum Electrodynamics (QED)~\cite{Landau:1965ksp,Migdal:1956tc}. Jet quenching consequently has complex dependence on the initial spatial configuration and the subsequent space--time evolution of the combined jet--QGP system. Systematic variation of the initial spatial configuration and evolution of the system therefore provides a valuable tool to elucidate the mechanisms underlying jet quenching, which is accomplished in practice by varying collision energy, collision geometry, and size of the colliding nuclei~\cite{Cunqueiro:2021wls,Apolinario:2022vzg,He:2024xtk}.

Significant jet quenching effects have been observed in large collision systems, Au+Au and Pb+Pb~\cite{Cunqueiro:2021wls,Apolinario:2022vzg}. In contrast, to date no clear evidence of jet quenching has been found in small collision systems, \pA\ and \pp~\cite{ALICE:2012mj,ATLAS:2014cpa,CMS:2016xef,STAR:2024nwm,PHENIX:2023dxl,Perepelitsa:2024eik}; limits on the magnitude of medium-induced energy loss due to quenching have  been reported for \pPb\ collisions at the LHC~\cite{ALICE:2017svf,ATLAS:2022iyq,CMS:2025jbv}. Further progress in varying system size to elucidate jet quenching mechanisms requires measurements in collisions of higher-mass projectiles, in which jet quenching signals may be measurable with sufficient precision to be discriminated from the effects of other processes. There has recently been a focus on the measurements in \OO\ collisions for this purpose~\cite{CMS:2025bta,STAR:2026nfy}. Jet quenching measurements for collisions of intermediate-mass nuclei, $A\sim$ 100, therefore provide a valuable complement to fill the gap between O+O ($A = 16$) and \AuAu\ and \PbPb\ ($A\sim200$) measurements.

In this paper, the STAR experiment at RHIC reports the first jet quenching measurements in $^{96}$Zr+$^{96}$Zr and $^{96}$Ru+$^{96}$Ru collisions at the center-of-mass energy per nucleon-nucleon pair (\sqrtsNN) of 200 GeV. Collisions of \RuRu\ and \ZrZr\ were recorded during the RHIC ``Isobar Run'' in 2018, each with an integrated luminosity of approximately 1.7 fb$^{-1}$. These datasets have been explored extensively to search for evidence of the Chiral Magnetic Effect, exploiting the different charge numbers of the Ru and Zr nuclei while maintaining the same mass number ($A=96$)~\cite{STAR:2021mii}. They also provide a unique opportunity for studying the system-size dependence of jet quenching as the radii of Ru and Zr are about 20\% smaller than that of Au and about a factor 1.8 larger than that of Oxygen. The difference in nuclear charge between Ru and Zr nuclei is expected to have a negligible impact on the QCD-driven jet-quenching phenomenon; this expectation is verified explicitly below.

The analysis reports the semi-inclusive yield of charged-particle jets recoiling from a high transverse momentum (high \pT) charged-hadron trigger~\cite{ALICE:2015mdb,STAR:2017hhs}. This observable is amenable to a fully data-driven statistical approach to mitigate the complex jet measurement background in heavy-ion collisions. It enables jet quenching measurements over broad phase space, including low jet \pT\ and large resolution parameter ($R$), where jet quenching effects are sizable~\cite{ALICE:2015mdb,STAR:2017hhs,ALICE:2019qyj,ALICE:2023jye,ALICE:2023qve,STAR:2023ksv,STAR:2023pal,STAR:2025yhg} and unique phenomena arising from the response of the QGP to jet excitation have been observed~\cite{ALICE:2023qve,STAR:2025yhg,Jing:2025bwi}. Uncorrelated background yield is corrected using event mixing~\cite{STAR:2017hhs,STAR:2023ksv,STAR:2023pal}. The recoil jet distributions are fully corrected for detector and background effects via unfolding. Jet quenching is quantified by comparing trigger-normalized recoil yields from central and peripheral collisions, and from jets with different $R$. The results are compared to similar measurements in \AuAu~\cite{STAR:2023ksv,STAR:2023pal} and \OO\ collisions~\cite{STAR:2026nfy}.

\section{Detector and dataset}
\label{sect:det}
The STAR experiment at RHIC is a general-purpose collider detector, with multiple subsystems for the measurement of hadrons, photons, electrons, and jets~\cite{STAR:2002eio}. The STAR detector provides precise charged-particle tracking and identification using a large Time Projection Chamber (TPC)~\cite{Anderson:2003ur} and the Time of Flight detector (TOF)~\cite{Llope:2005yw} in a 0.5 Tesla magnetic field.

This analysis uses data recorded for \RuRu\ and \ZrZr\ collisions at $\sqrtsNN=200$ GeV with a minimum bias (MB) trigger that requires coincident signals in the forward Vertex Position Detectors (VPD) with pseudo-rapidity acceptance $4.24<|\eta|<5.10$~\cite{Llope:2014nva}, and in Zero-Degree Calorimeters (ZDC) which detect neutrons at beam rapidity~\cite{Adler:2003sp}. The MB trigger samples 78.7\% and 77.5\% of the total inelastic cross section for \RuRu\ and \ZrZr\ collisions, respectively, estimated by the Glauber model~\cite{Miller:2007ri}. The slightly larger trigger efficiency for \RuRu\ collisions arises from larger particle multiplicity than that for \ZrZr\ collisions~\cite{STAR:2021mii}.

Charged-particle tracks from signal hits in the TPC are constructed using the Kalman Filter algorithm~\cite{Kalman:1960mft,Fruhwirth:1987fm} and are denoted global tracks. The primary collision vertex is constructed from global tracks. Events are accepted for further analysis for $-35<\zvtxTPC<25$ cm, where \zvtxTPC\ is the position of the primary vertex in the beam direction relative to the center of the STAR detector. The asymmetric cut accounts for an offset in the peak position of the \zvtxTPC\ distribution due to beam conditions. Event pileup is suppressed by the requirement $|\zvtxTPC-\zvtxVPD|<5$ cm, where \zvtxVPD\ is determined using the VPD signal; VPD is a fast detector, whose signal is less susceptible to pileup than that of the TPC.

The global track momentum is refit by including the primary vertex to improve the \pT\ resolution. The resulting tracks, referred to as primary tracks, are used in jet reconstruction. The contribution of displaced weak-decay vertices is suppressed by requiring the distance of closest approach to the primary vertex of the corresponding global track (gDCA) to be less than 1 cm. The residual population of displaced-vertex tracks can erroneously be assigned high \pT\ by the primary-track fit if the primary vertex lies outside the helix of the track trajectory projected onto the transverse plane, generating a background to true high-\pT\ tracks whose production rate is low. To suppress this contribution, an additional cut is applied to the signed DCA, i.e., sDCA $<0.5$ cm for positive tracks and sDCA $>-0.5$ cm for negative tracks. The sign is defined relative to the global-track curvature in the magnetic field, such that tracks bending toward the primary vertex have positive sDCA. The following additional criteria are imposed on primary tracks: a minimum of 15 TPC hits used in track reconstruction; the ratio of the number of TPC hits to the possible number of hits along the track trajectory greater than 0.52; $0.2<\pT<25$ GeV/$c$; and $|\eta|<1$. 

Events are classified offline in percentile bins of the multiplicity-ordered charged-particle distribution measured using primary tracks within $|\eta|<0.5$ (``centrality''), corrected for the online MB trigger inefficiency using the Glauber model~\cite{Miller:2007ri,STAR:2021mii}. Events are analyzed in two centrality classes corresponding to percentile bins 0--10\% (``central'') and 60--80\% (``peripheral''), where 0\% indicates the largest multiplicity. After all event selection cuts, the \ZrZr\ dataset has 261 million central and 368 million peripheral events, while the \RuRu\ dataset has 251 million central and 349 million peripheral events. The \RuRu\ and \ZrZr\ datasets are analyzed separately, and then combined for the physics results reported below.

The TPC tracking performance is evaluated by simulating single pions, kaons, protons and their anti-particles, propagating them through a GEANT 3 model of the full STAR detector, and embedding these detector-level tracks into real events that are processed the same way as real data. Tracking performance is very similar in the two collision systems. The primary tracking efficiency is about 65\% for central and 70\% for peripheral collisions at $\pT=0.2$ GeV/$c$,  83\% and 86\% respectively at 3 GeV/$c$, and 86\% and 89\% respectively for $\pT>10$ GeV/$c$. The track \pT\ resolution is about 0.8\% at $\pT=0.2$ GeV/$c$, 2.0\% at 3 GeV/$c$, and 6.3\% at 10 GeV/$c$, with negligible centrality dependence. 

\section{Observables}
For the semi-inclusive measurements, events are selected additionally by the requirement of a high-\pT\ charged hadron (``trigger''), chosen such that their ensemble-averaged \pT\ distribution corresponds to that of inclusive charged-hadron production. Jet reconstruction is carried out on these selected events (Sect.~\ref{sect:jet_reco}), and the number of jets in the recoil acceptance, which is azimuthally opposite to the trigger direction, is counted. 

The reported observable  is the \pT-differential distribution of the trigger-normalized recoil jet yield,
\begin{equation}
\YhJet(\pTreco) = \frac{1}{\Ntrig}\cdot
\frac{\mathrm{d}\Njetch}{\mathrm{d}\etajetch\mathrm{d}\pTreco},
\label{eq:yield}
\end{equation} 
\noindent
where \Ntrig\ is the observed number of triggers, \etajetch\ and \pTreco\ are the pseudo-rapidity and reconstructed transverse momentum of a recoil jet, and \Njetch\ is the number of recoiling jets in the dataset. \YhJet(\pTreco) requires correction for background and detector effects, described below, and the corrected distributions are denoted \YtildehJet(\pTjetch), where \pTjetch\ denotes corrected recoil jet \pT. The observable is semi-inclusive, due to the counting procedure for the trigger hadron and recoil jets, and  \YtildehJet\ is therefore equivalent to the ratio of two hard cross sections: coincidence production of a trigger hadron and recoiling jet within the acceptance, and inclusive charged hadron production~\cite{ALICE:2015mdb,STAR:2017hhs}.

Yield suppression due to jet quenching is quantified by \ICP, the ratio of \YtildehJet(\pTjetch) in central and peripheral collisions,
\begin{equation} 
    \ICP(\pTjetch) = \frac{\YtildehJet_{\mathrm{Central}}}{\YtildehJet_{\mathrm{Peripheral}}}.
\label{eq:Icp}
\end{equation}
\noindent
Here, \ICP\ is used since a corresponding \pp\ reference for the trigger and recoil-jet selections used here is not available, precluding a measurement of \IAA.

Broadening of internal jet structure due to jet quenching is quantified by \Rbrtwofive, the ratio of \YtildehJet(\pTjetch) for jets reconstructed in the same centrality bin with $R=0.2$ and 0.5,
\begin{equation} 
    \Rbrtwofive(\pTjetch)=\frac{\YtildehJet_{R=0.2}}{\YtildehJet_{R=0.5}}.
\label{eq:Rtwofive}
\end{equation}

\section{Analysis}

The analysis procedure follows closely that described in Ref.~\cite{STAR:2017hhs}. After trigger selection and jet reconstruction, uncorrelated background jet yield is subtracted using event mixing, and corrections for \pT-smearing due to detector effects and residual background fluctuations are then applied by unfolding.

\subsection{Trigger hadrons and event selection}

Charged-particle tracks with $7<\pTtrig <25$ GeV/$c$ and $|\eta|<1$ are selected as trigger hadrons. For events with multiple tracks satisfying these conditions, one track is selected at random to ensure that the trigger hadron distribution is the same as that of inclusive charged-hadron production~\cite{ALICE:2015mdb,STAR:2017hhs}. For the 0--10\% centrality class, these multi-trigger events constitute about 1\% of the total events containing at least one trigger hadron. The lower bound on trigger \pT\ is chosen as 7 GeV/$c$ rather than 9 GeV/$c$, as in previous related work~\cite{STAR:2017hhs, STAR:2023ksv, STAR:2023pal}, to ensure sufficient statistical precision in peripheral collisions. The combined \RuRu\ and \ZrZr\ datasets contain 852k triggers for central and 87k triggers for peripheral collisions.

\subsection{Jet reconstruction}
\label{sect:jet_reco}

Jet reconstruction using all accepted charged-particle tracks is carried out twice. The first jet reconstruction pass utilizes the \kT\ algorithm~\cite{Catani:1993hr} with the $E$-scheme and $R=0.4$, implemented in the FastJet package~\cite{Cacciari:2005hq,Cacciari:2011ma}. Jet candidates are accepted if their centroid lies within $|\etajetch|<1-R$ to avoid partially reconstructed jets. An event-wise estimate of the background is determined by the median of the ratio $\rho=\pTraw/\Ajet$ for all first-pass jet candidates except the two with largest \pTraw, where \Ajet\ is the jet area~\cite{Cacciari:2007fd}.

The second jet reconstruction pass utilizes the \antikT\ algorithm~\cite{Cacciari:2008gp} with the $E$-scheme, likewise implemented in FastJet, for $R=0.2,0.3,0.4,$ and 0.5. Accepted jet candidates from the second pass are also required to have centroid $|\etajetch|<1-R$, and to have a value of \Ajet\ greater than 0.05, 0.20, 0.35, or 0.65, for $R=0.2,0.3,0.4$ and 0.5, respectively, to suppress unphysical background jets~\cite{STAR:2017hhs}. The value \pTraw\ of each accepted jet candidate is then adjusted according to~\cite{Cacciari:2010te}

\begin{equation}
\pTreco = \pTraw - \rho\cdot\Ajet. 
\label{rho}
\end{equation}

\noindent
The recoil acceptance is $3\pi/4\le\dphi\le\pi$, where $\dphi\in[0,\pi]$ is the azimuthal separation between the trigger direction and the jet centroid. 

\subsection{Mixed-event correction} 
\label{ME process}

The recoil jet population includes correlated, physical jets arising from the same high-\Qsqr\ process as the trigger hadron, and uncorrelated jets arising from both the combinatorial combination of particles from soft (low \Qsqr\ interactions), and physical jets from other high-\Qsqr\ processes (multiple partonic interactions, or MPIs). The rate of recoil jets originating from MPIs is negligible compared to the correlated yield at RHIC energies~\cite{STAR:2017hhs}.

To correct for the yield of uncorrelated combinatorial jets, the analysis utilizes the event mixing procedure~\cite{STAR:2017hhs,STAR:2023ksv,STAR:2023pal} to remove inter-particle correlations and reconstruct combinatorial jets. Mixed events (ME) are created from tracks in real data events (``Same Events'', or SE). Only tracks with $\pT<5$ GeV/$c$ are used to avoid potential influence of signal jets. The SE population is classified in 8 bins in \Nch, 20 bins in \zvtxTPC, and 4 bins in event plane (EP) orientation, where \Nch\ is the number of primary tracks in the azimuthal quadrants transverse to the trigger direction. A set of ME events is constructed for each SE bin, with each ME event having at most one track from a given SE. The ME distribution in \Nch\ corresponds to that for the SE population. ME events from all bins are then combined to form the ME population for further analysis. The same procedure used to construct \YhJet(\pTreco) in SE events is applied to ME population, with two differences: (i) a random direction in $\varphi$ is chosen as the trigger direction, and (ii) no jets are removed in the evaluation of $\rho$. A good agreement in the $\rho$ distribution is seen between SE and ME population.

\begin{figure*}[tb]
\centering
\includegraphics[width=0.8\textwidth]{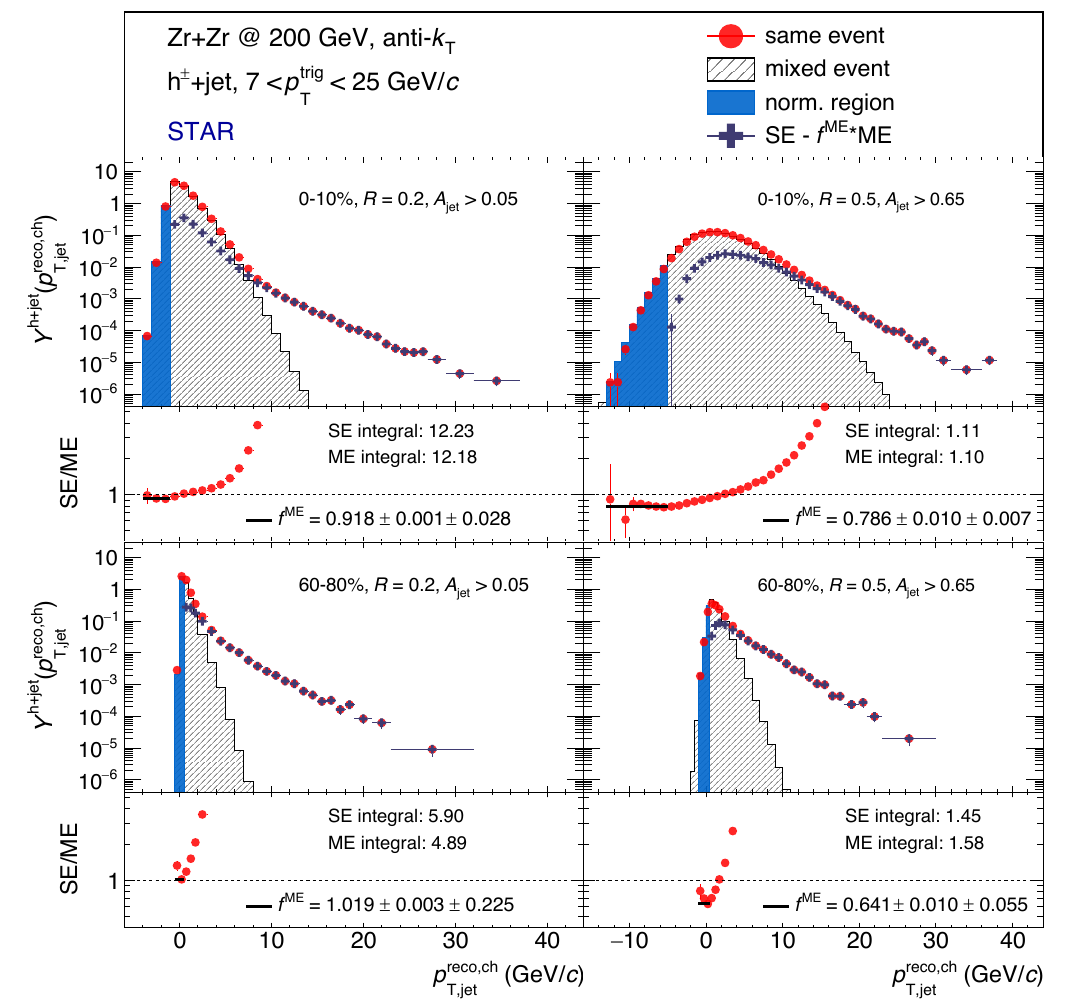}
\caption{Distributions of \YhJet(\pTreco) in central (upper panels) and peripheral (lower panels) \ZrZr\ collisions for $R=0.2$ (left panels) and $R=0.5$ (right panels). Upper sub-panels: SE (red points) and ME (shaded histogram) distributions; ME normalization region (blue histogram);  and difference distribution (black crosses). Error bars are statistical only. Lower sub-panels: ratio of \YhJet(\pTreco) for SE/ME. Solid horizontal lines indicate fits to the SE/ME ratio in the normalization region. Resulting \fME\ values, along with their statistical (first) and systematic (second) uncertainties, are also listed.}
\label{SE ME Jet}
\end{figure*}

Figure~\ref{SE ME Jet} shows SE and ME distributions of \YhJet(\pTreco) for central and peripheral \ZrZr\ collisions, for $R=0.2$ and 0.5. Distributions for \RuRu\ collisions are quantitatively similar. Since no jet candidates are rejected in the analysis, the SE distributions contain a mixture of physical jets that are correlated with the trigger and uncorrelated background jets. 

For central collisions, the SE distributions exhibit a broad peak centered at $\pTreco\sim0$ and a prominent tail at large positive \pTreco. Since $\rho$ is the median \pT-density in each event, about half of the jet candidate population have $\pTreco<0$ (Eq.~\ref{rho}). The ME distributions exhibit a similar broad peak at $\pTreco\sim0$, though without the tail at large positive \pTreco\ of the SE distributions. In the region $\pTreco<0$, the yield ratios SE/ME vary by a few percent, as illustrated in the lower sub-panels of Fig.~\ref{SE ME Jet}, while the individual yields vary by five orders of magnitude. The SE and ME distributions therefore have very similar shape in that region, indicating dominance of the SE yield by combinatorial jets and validating the ME distribution as an estimate of the combinatorial jet contribution~\cite{ALICE:2015mdb,STAR:2017hhs}. Additionally, the integrals over \pTreco\ of the SE and ME distributions, as listed in lower sub-panels, are consistent within less than 1\%, indicating that the total number of \antikT\ jet candidates in the acceptance is conserved, independent of the \pTreco\ distribution~\cite{ALICE:2015mdb,STAR:2017hhs}. The combinatorial yield in the SE population is therefore taken as the scaled ME distribution, where the scaling factor \fME\ is used to account empirically for the reduced acceptance for combinatorial jets in the SE population due to the presence of true correlated jets~\cite{ALICE:2015mdb,STAR:2017hhs}. \fME\ is determined by fitting the SE/ME yield ratio in the negative \pTreco\ region (denoted ``normalization region" in Fig.~\ref{SE ME Jet}), as illustrated by horizontal solid lines in the lower sub-panels of Fig.~\ref{SE ME Jet}. The difference distribution between SE distribution and scaled ME distribution, SE $-$ \fME$*$ME, represents the physical recoil jet yield correlated with the trigger (Fig.~\ref{SE ME Jet}), whose correction procedure is discussed later.

For peripheral collisions, qualitatively similar features are observed. However, the combinatorial jet distribution in $\pTreco<0$ is narrower, corresponding to a reduced uncorrelated background. In addition, the \pTreco\ integrals of SE and ME differ by a larger factor, and the yield ratio SE/ME does not exhibit a significant range that is independent of \pTreco. These differences arise because of the much lower multiplicity in peripheral collisions, such that the acceptance is not fully occupied by reconstructed jets~\cite{ALICE:2015mdb,STAR:2017hhs}. The same ME scaling and subtraction procedure is nevertheless applied for consistency, with the resulting uncertainties incorporated in the total systematic uncertainty. Since the relative background contribution is much smaller in peripheral collisions, this has only minor impact on the corrected results.

\subsection{Unfolding}
The event-wise \pT-correction (Eq.~\ref{rho}) is approximate, with residual background fluctuations and detector effects that are corrected by unfolding~\cite{Brenner:2019lmf}. The response matrix for unfolding, \Rtot(\pTjetch,\pTreco), encodes mapping from truth to reconstructed recoil jet \pT:
\begin{equation}
\YtildehJet(\pTjetch) = \tilde{R}_{\mathrm{tot}}^{-1}(\pTjetch,\pTreco) \otimes \YhJet(\pTreco),
\label{unfoldmatrix}
\end{equation}
where $\tilde{R}_{\mathrm{tot}}^{-1}(\pTjetch,\pTreco)$ denotes the regularized inversion of \Rtot(\pTjetch,\pTreco). Here, \Rtot(\pTjetch,\pTreco) can be factorized as the product of two components, the background-fluctuation response matrix \Rbkg(\pTdet,\pTreco) and the detector response matrix \Rdet(\pTpart,\pTdet), where \pTpart\ and \pTdet\ are jet \pT\ at the particle and detector levels. In practice, \pTpart\ is equal to \pTjetch.

Particle-level events are simulated using PYTHIA 6.4 STAR tune~\cite{Adkins:2015ccl}. Detector-level events are obtained by fast simulation, applying the tracking efficiency and \pT\ resolution to particle-level tracks. Reconstructed particle-level and detector-level jets in the corresponding  events are matched if they have the smallest angular  centroid separation, $\sqrt{\Delta \eta^2 + \Delta \phi^2}$, of all match candidates, with the separation less than $R$, and with $\pTdet/\pTpart>0.15$. The matrix \Rdet(\pTpart,\pTdet) is constructed from matched particle-level and detector-level jet pairs. Unmatched jets are also corrected for through the jet-finding efficiency and the fake-jet rate. The former is the fraction of all particle-level jets that are matched, while the latter is the fraction of all detector-level jets that are unmatched. The jet-finding efficiency for $R=0.2$ is 91\% for central and 93\% for peripheral collisions at $\pTpart=5$ GeV/$c$, and 97\% for central  and 98\% for peripheral collisions at $16<\pTpart<20$ GeV/$c$, and is slightly larger for larger $R$. The fake jet rate for $R=0.5$ is about 3\% at $\pTdet=5$ GeV/$c$ and 1\% at $\pTdet=20$ GeV/$c$ in both central and peripheral collisions, and is smaller for smaller $R$. 

The background response matrix \Rbkg(\pTdet,\pTreco) is determined by embedding, with each jet represented as a single four vector (``single particle'' or SP embedding~\cite{ALICE:2015mdb,STAR:2017hhs}). The SP approach to measuring background fluctuation effects is preferred since the resulting response matrix has been shown to be insensitive to the specific fragmentation model applied, as expected for infrared and collinear-safe (IRC-safe) jet reconstruction, and the matching of SP ``jets'' is unambiguous.

The corrected jet yield distribution as a function of \pTjetch\ is determined by iterative Bayesian unfolding~\cite{DAgostini:1994fjx} implemented in the RooUnfold package ~\cite{Brenner:2019lmf,Roounf}. Matrix elements of \Rtot(\pTjetch,\pTreco) with \pTreco\ or \pTjetch\ less than 4 GeV/$c$ are set to zero, and the resulting missing jets are included in the jet-finding efficiency and the fake-jet rate. The optimal prior, determined by the consistency of the back-folded and measured distributions, is the spectrum calculated using PYTHIA 6.4 STAR tune, scaled by a \pT-dependent factor which hardens the distribution. The back-folded distribution is obtained by convoluting the unfolded distribution with \Rtot(\pTjetch,\pTreco), and variation in the \pT-dependent factor contributes to the systematic uncertainty. The optimum number of unfolding iterations is 5 for central and 3 for peripheral collisions, determined as the lowest number of iterations for which variations in unfolded spectra in successive iterations are comparable in magnitude with the statistical errors.

A closure test is performed using events simulated by PYTHIA 6.4 STAR tune with comparable statistical precision to the data, applying detector effects and background fluctuations, and unfolding using the same procedures as those for real data. Comparison of the corrected data with the input particle-level distribution shows good agreement within statistical errors, validating the full correction procedure.

\subsection{Systematic uncertainties}

Systematic uncertainties are assessed by varying the analysis components as follows: tracking efficiency ($4\%$ absolute uncertainty); upper \pT\ limit on tracks used in event mixing (4 and 7 GeV/$c$ vs. default 5 GeV/$c$); fitting range for $f^{\rm ME}$ ($\pm 1$ GeV/$c$ for central; $\pm 0.5$ GeV/$c$ for peripheral collisions); tracking efficiency for weak-decay daughters; jet \pT\ lower limit in \Rtot(\pTjetch,\pTreco) ($\pm1$ GeV/$c$); prior distribution shape (steeper or flatter than default); and unfolding iteration number ($\pm1$ relative to optimum value). 

Correlation of trigger hadrons with EP orientation can induce a bias in the uncorrelated background distribution in the recoil jet acceptance. However, such correlations were found to have negligible effect on corrected recoil jet distributions in central Au+Au collisions~\cite{STAR:2017hhs}. Insofar as the relative magnitude of uncorrelated background and the effects of jet quenching are both smaller in \RuRu\ and \ZrZr\ collisions, such correlations are likewise expected to have negligible effect in this analysis and were not evaluated explicitly.

The total systematic uncertainty is the quadrature sum of all component uncertainties. For $6<p_{\mathrm{T,jet}}^{\mathrm{ch}}<8$ GeV/$c$, the dominant uncertainties in \ICP\ for $R=0.2$ arise from the lower jet \pT\ limit on the response matrix (6.6$\%$), track \pT\ limit for event mixing (4.1$\%$), and unfolding (2.8$\%$), with a total uncertainty of 8.4$\%$, while for $19<p_{\mathrm{T,jet}}^{\mathrm{ch}}<24$ GeV/$c$ the dominant uncertainties are unfolding (3.2$\%$) and tracking efficiency (2.6$\%$), with a total uncertainty of 4.2$\%$.

\section{Results}

\begin{figure}[hbtp]
	\centering
    \includegraphics[width=0.49\textwidth]{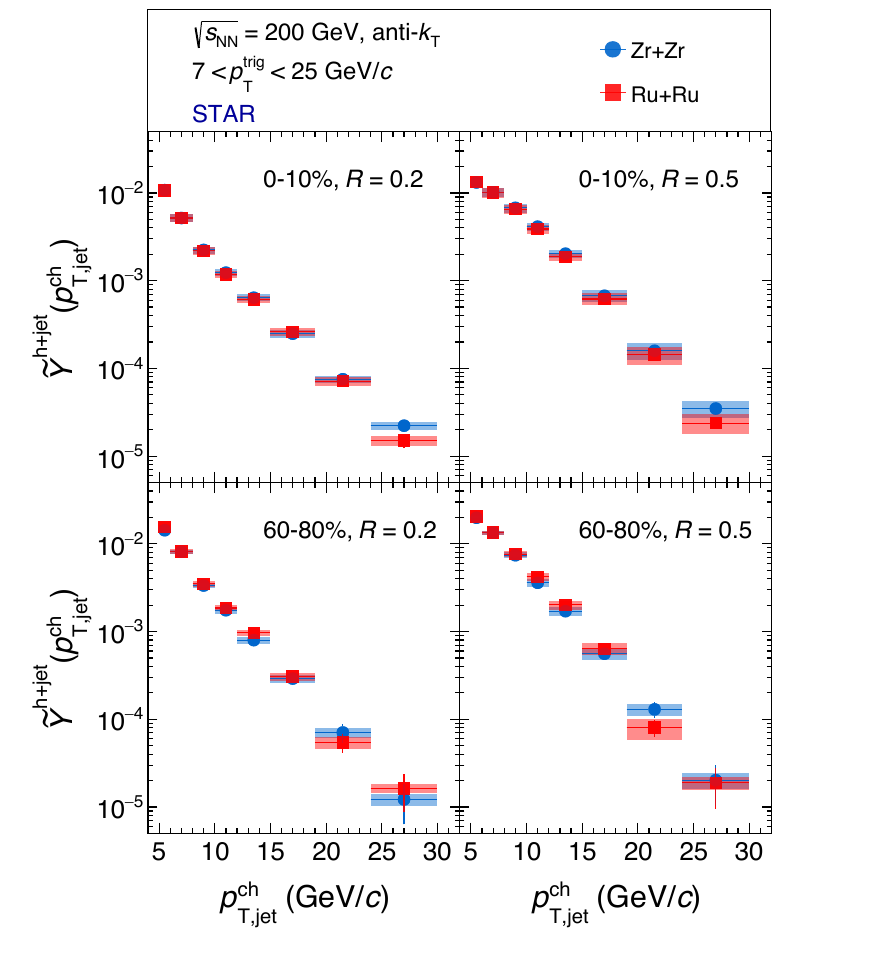}
\caption{Distributions of  \YtildehJet(\pTjetch) for \RuRu\ and \ZrZr\ collisions at $\sqrtsNN= 200$ GeV, for $R= 0.2$ (left) and 0.5 (right). Upper panels: central collisions; lower panels: peripheral collisions. Statistical errors and systematic uncertainties are indicated by error bars and shaded boxes, respectively.}
\label{fully correct spectra}
\end{figure}

Figure~\ref{fully correct spectra} shows the \YtildehJet(\pTjetch) distributions in central and peripheral \RuRu\ and \ZrZr\ collisions at $\sqrtsNN=200$ GeV for $R=0.2$ and 0.5. Similar distributions for $R=0.3$ and 0.4 are shown in the Appendix. The distributions for \RuRu\ and \ZrZr\ collisions are consistent, and are combined in the following using the inverse of the statistical variance as weights; the combined distributions are shown in the Appendix.

\begin{figure}[htb]
\centering
\includegraphics[width=0.45\textwidth]{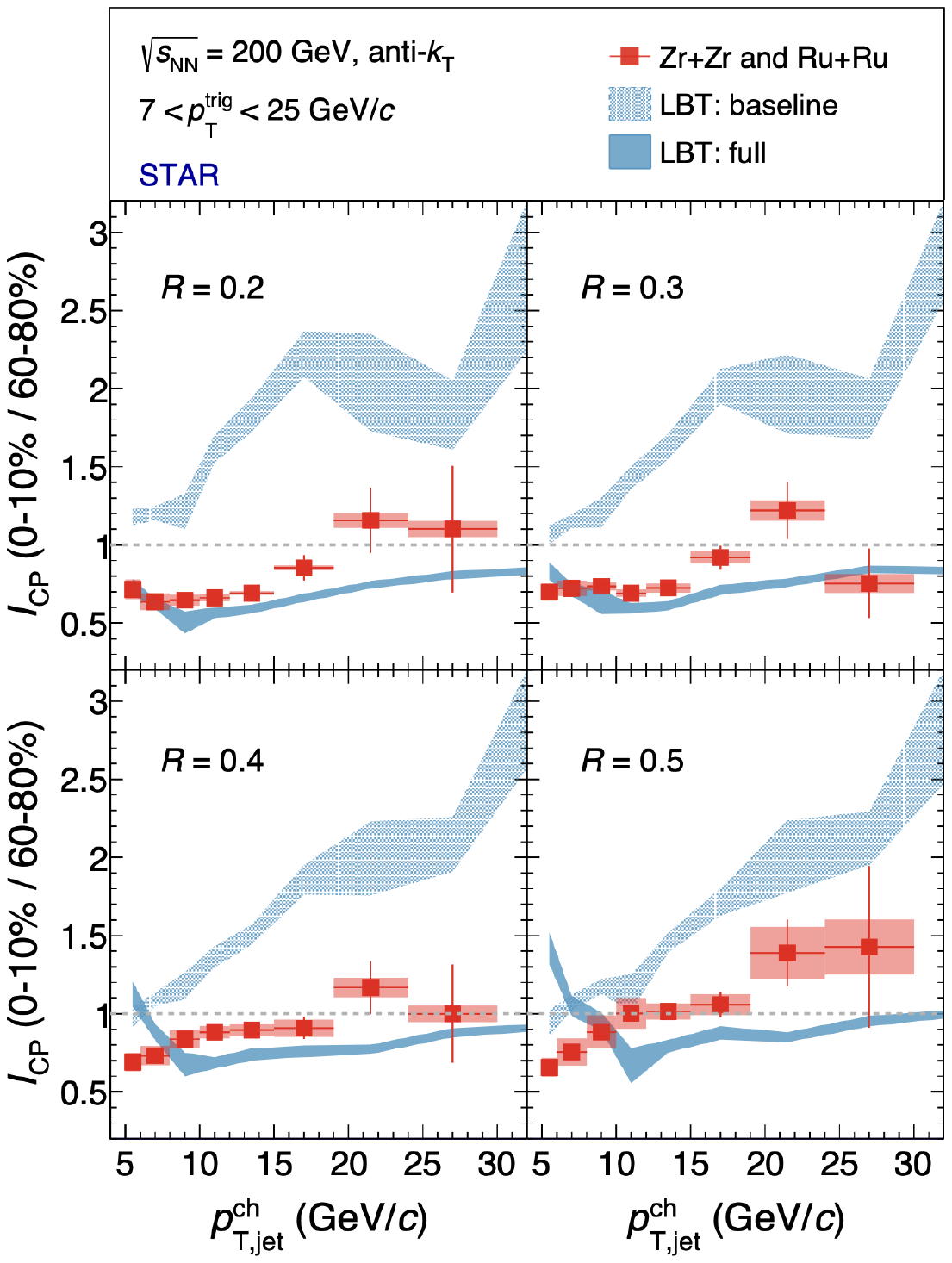}
\caption{Distributions of \ICP\ for combined \RuRu\ and \ZrZr\ data at $\sqrtsNN= 200$ GeV, with $R=0.2$, 0.3, 0.4 and 0.5. Statistical errors and systematic uncertainties are indicated by error bars and shaded boxes. Shaded bands show corresponding LBT model calculations incorporating energy loss for jets generating observed trigger hadrons, with (``LBT:full") and without (``LBT:baseline") energy loss for recoil jets~\cite{Dang:2026ezw,privatecommLBT}. The width of the band indicates the statistical error of the calculation.}
\label{fully correct merged Icp}
\end{figure}

Figure~\ref{fully correct merged Icp} shows the distribution of  \ICP\ (Eq.~\ref{eq:Icp}) for all $R$ values. The distributions exhibit a common trend, with $\ICP <1$ at low \pTjetch\ region, gradually increasing and approaching unity at higher \pTjetch. The significant suppression in yield  at low \pTjetch\ for central relative to peripheral collisions provides direct evidence of partonic energy loss due to jet interactions in the QGP. 

The rise in \ICP\ with increasing \pTjetch\ is similar to that seen in h+jet measurements of Pb+Pb collisions at \sqrtsNN\ = 5.02 TeV, with the \pTjetch\ scale where the rise occurs increases with increasing \pTtrig~\cite{ALICE:2023qve}. Theoretical considerations indicate that high-\pT\ hadron production in heavy-ion collisions is subject to a geometric bias, whereby the locus of generation of {\it observed} hadrons is biased towards the surface of the QGP, with their trajectory headed outward, due to the interplay of the shapes of the jet production spectrum and fragmentation function, and jet quenching~\cite{Baier:2002tc,Dainese:2004te,Renk:2006nd,Zhang:2007ja}. These considerations motivate the choice of a high-\pT\ hadron trigger in semi-inclusive analyses, which can maximize the recoil jet path-length in the QGP~\cite{ALICE:2015mdb,STAR:2017hhs}. 

Recent studies based on the Linear Boltzmann Transport (LBT) model~\cite{Cao:2016gvr,Luo:2023nsi} show that such surface bias effects may be modified in practice due to finite energy loss of the jets generating observed trigger hadrons~\cite{He:2024rcv}, resulting in an increase in the \ICP\ observable even for significant energy loss in the recoil jet population. Figure~\ref{fully correct merged Icp} shows results of such LBT calculations at the particle level~\cite{Dang:2026ezw,privatecommLBT}, using a value for the strong coupling $\alpha_{s}$ that was previously tuned to describe jet quenching measurements in \AuAu\ and \PbPb\ collisions~\cite{Zhang:2022ctd}. The distribution labeled ``LBT: baseline" accounts for partonic energy loss of the parent jets that generate observed trigger hadrons but not of the recoil jets. It therefore does not represent the absence of medium effects; rather, it serves as a reference for isolating the quenching effect on the recoil jet population. The distribution labeled ``LBT: full" includes jet quenching for both trigger hadrons and the recoil population. The value of  \ICP\  for the LBT baseline calculation lies well above this measurement for \RuRu\ and \ZrZr\ collisions, while that for the LBT full calculation lies below the current measurements above 10 GeV/$c$, though much closer to the data.

\begin{figure}[htb]
\centering
\includegraphics[width=0.5\textwidth]{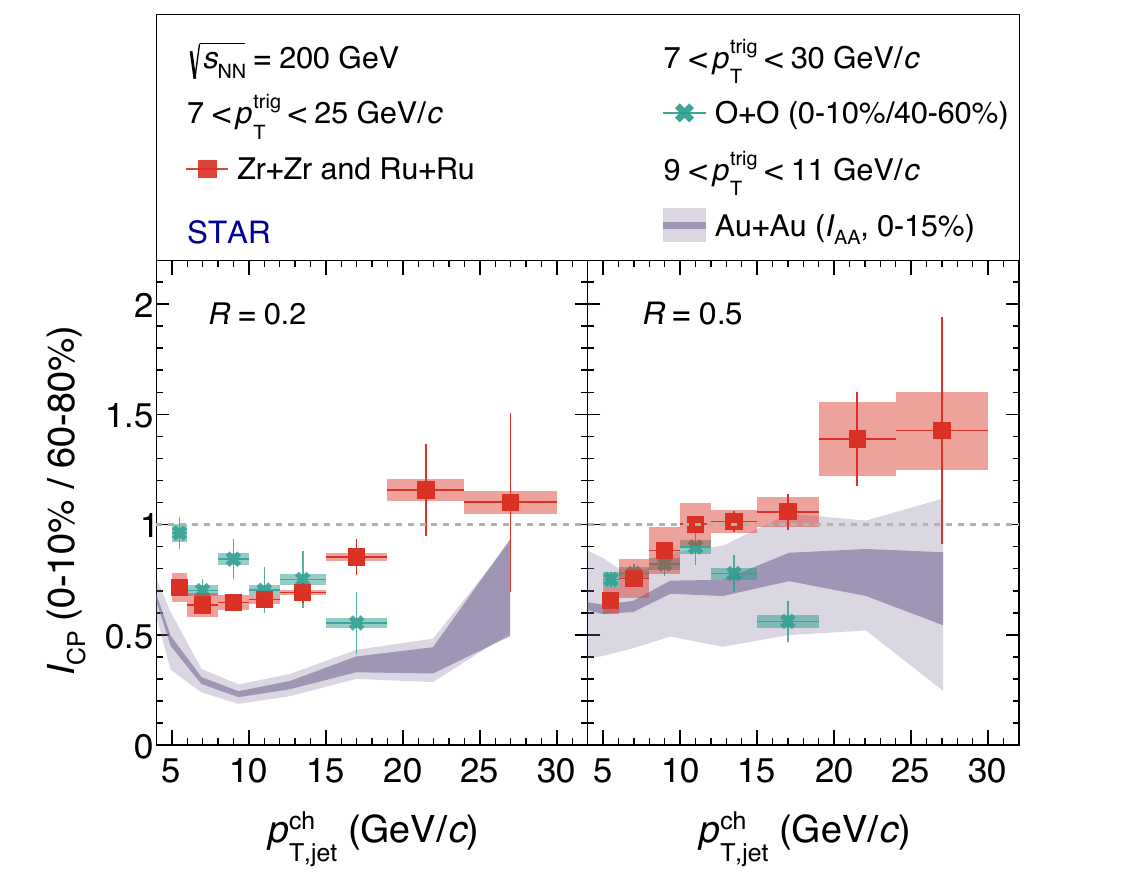}
\caption{Distributions of \ICP\ for combined \RuRu\ and \ZrZr\ data at $\sqrtsNN= 200$ GeV from Fig.~\ref{fully correct merged Icp} compared to \IAA\ for central \AuAu\ collisions for \pizero\ triggers with $9<\pTtrig <11$ GeV/$c$~\cite{STAR:2023ksv,STAR:2023pal}, and \ICP\ for high-EA \OO\ collisions for charged-hadron triggers with $7<\pTtrig <30$ GeV/$c$~\cite{STAR:2026nfy}. }
\label{fully correct Icp R2 R5}
\end{figure}
Figure~\ref{fully correct Icp R2 R5} compares \ICP\ for combined \RuRu\ and \ZrZr\ data to distributions of \IAA\ (i.e. normalized using \pp\ rather than peripheral \aaa\ collision data) from a semi-inclusive recoil-jet measurement in  0-15$\%$ central \AuAu\ collisions for \pizero\ triggers with $9<\pTtrig <11$ GeV/$c$~\cite{STAR:2023ksv, STAR:2023pal}, and to \ICP\ of high event activity (EA) \OO\ collisions for charged-hadron triggers with $7<\pTtrig <30$ GeV/$c$~\cite{STAR:2026nfy}. While these measurements differ in centrality selection, trigger definition, and the use of \IAA\ versus \ICP, which precludes their direct quantitative comparison, they are nevertheless closely related and provide key benchmarks which constrain the modeling of jet-quenching across collision systems. For $R=0.2$, significantly larger yield suppression is observed for \AuAu\ than for the combined \RuRu\ and \ZrZr\ data, while the suppression in \OO\ collisions is similar in magnitude to that of the combined data. For $R=0.5$ the yield is less suppressed than for $R=0.2$ in \AuAu\ collisions, suggesting recovery of energy in the larger aperture. However, the larger uncertainties of the \AuAu\ collision data preclude definitive comparison with the combined \RuRu\ and \ZrZr\ data. 

The conversion of yield suppression, measured by \ICP\ and \IAA, to recoil-jet energy loss must take into account the trigger hadron energy loss (Fig.~\ref{fully correct merged Icp}), which likely differs between collision systems. It is worth noting that previous STAR measurements~\cite{STAR:2016jdz} indicate that the fragmentation patterns of partons producing $\pi^{0}$ triggers with $12<\pTtrig<20$~GeV/$c$ are consistent between \pp\ and central \AuAu\ collisions. Calculations which include these effects are needed to quantify system size dependence of partonic energy loss in these different collision systems. 

\begin{figure}[htbp]
\centering
\includegraphics[width=0.45\textwidth]{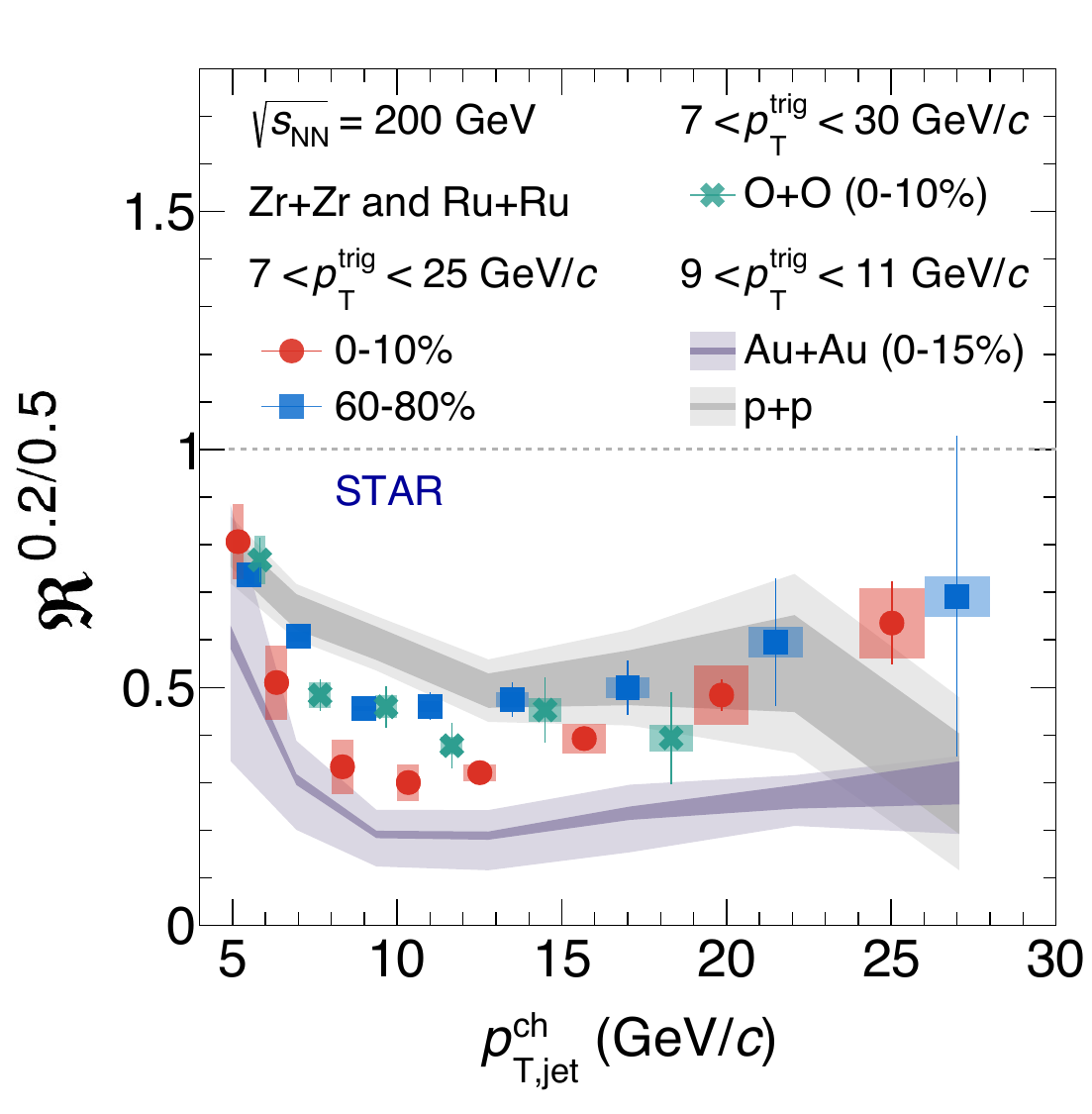}
\caption{Distributions of \Rbrtwofive\ for central and peripheral \RuRu\ and \ZrZr\ collisions at $\sqrtsNN=200$ GeV. Also shown are \Rbrtwofive\ for central \OO\ collisions with a charged-hadron trigger~\cite{STAR:2026nfy}, and for \pp\ and central \AuAu\ collisions  with a \pizero\ trigger~\cite{STAR:2023ksv, STAR:2023pal}.}
\label{Rtwofive}
\end{figure}

Figure~\ref{Rtwofive} shows the distribution of \Rbrtwofive\  (Eq.~\ref{eq:Rtwofive}), which probes medium-induced broadening of internal jet structure, for the combined data from Fig.~\ref{fully correct spectra}. Similar distributions for \Rbrthreefive\ and \Rbrfourfive\ are included in the Appendix. The effect of trigger hadron energy loss is predicted by LBT to largely cancel in the \Rbrtwofive\ ratio~\cite{He:2024rcv}. The value of \Rbrtwofive\ is markedly smaller in central than peripheral collisions for $\pTjetch>8$ GeV/$c$, indicating significant medium-induced broadening of the jet energy profile for angles less than 0.5 rad. Also shown are distributions of  \Rbrtwofive\ for 0--15\% \AuAu\ and \pp\ collisions with a \pizero\ trigger of 9--11 GeV/$c$~\cite{STAR:2023ksv, STAR:2023pal} and for 0--10\% high-EA \OO\ collisions with a charged-hadron trigger of 7--30 GeV/$c$~\cite{STAR:2026nfy}. The peripheral collision data from this measurement are largely consistent with those from \pp\ collisions, indicating that the difference in trigger \pTtrig\ selection in the two analyses does not have a strong influence on this observable. The \Rbrtwofive\ ratio in central collisions decreases sequentially with increasing system size, from \OO\ to \RuRu/\ZrZr\ to \AuAu, indicating stronger medium-induced modification of jet structure in larger systems. 

Figure~\ref{fully correct Icp R2 R5} indicates recoil jet yield suppression is smaller in combined \RuRu\ and \ZrZr\ data than in central \AuAu\ collisions. Figure~\ref{Rtwofive} likewise shows greater medium-induced jet broadening in a larger collision system. These comparisons provide new insight into the system-size dependence of jet quenching. Similar yield suppression and intra-jet broadening has been observed in h+jet measurements of \PbPb\ collisions at $\sqrtsNN=5.02$ TeV~\cite{ALICE:2023qve}, whose comparison to these data and those in Ref.~\cite{STAR:2023ksv,STAR:2023pal} will further elucidate the collision energy and QGP temperature dependence of jet quenching. Quantitative comparison of these data with detailed theoretical models of jet quenching requires a comprehensive Bayesian Inference analysis approach~\cite{JETSCAPE:2024cqe}.

\section{Summary}

The STAR Collaboration at RHIC reports the first measurement of semi-inclusive hadron-jet distributions in central and peripheral \RuRu\ and \ZrZr\ collisions at \sqrtsNN\ = 200 GeV, for jet resolution parameters between $R=0.2$ and 0.5. Recoil jet spectra from \RuRu\ and \ZrZr\ collisions are consistent, and their corrected datasets are combined. The measurements reveal clear medium-induced effects due to jet quenching: suppression of recoil jet yield and broadening of the transverse jet profile. Recoil yield suppression diminishes with increasing \pTjetch, consistent with model calculations of the effect of jet quenching on the trigger hadron spectrum. The observed suppression magnitude is less than that in central \AuAu\ collisions. Comparison of the medium-induced intra-jet broadening with that observed in \OO\ and \AuAu\ collisions likewise shows a distinct ordering of jet quenching effects with system size. These data provide novel and significant constraints on the system-size dependence of jet quenching phenomena, and correspondingly theoretical models incorporating spatial and temporal dependence of the physical mechanisms underlying jet quenching. 

\section{Acknowledgments}
We thank the RHIC Operations Group and SCDF at BNL, the NERSC Center at LBNL, and the Open Science Grid consortium for providing resources and support.  This work was supported in part by the Office of Nuclear Physics within the U.S. DOE Office of Science, the U.S. National Science Foundation, National Natural Science Foundation of China, Chinese Academy of Science, the Ministry of Science and Technology of China and the Chinese Ministry of Education, NSTC Taipei, the National Research Foundation of Korea, Czech Science Foundation and Ministry of Education, Youth and Sports of the Czech Republic, Hungarian National Research, Development and Innovation Office, New National Excellency Programme of the Hungarian Ministry of Human Capacities, Department of Atomic Energy and Department of Science and Technology of the Government of India, the National Science Centre and WUT ID-UB of Poland, German Bundesministerium f\"ur Bildung, Wissenschaft, Forschung and Technologie (BMBF), Helmholtz Association, Ministry of Education, Culture, Sports, Science, and Technology (MEXT), Japan Society for the Promotion of Science (JSPS), and Agencia Nacional de Investigacion y Desarrollo de Chile (ANID), Chile.

\section*{Appendix}
\label{sect:Appendix}

\begin{figure}[hbtp]
\centering
 \includegraphics[width=0.49\textwidth]{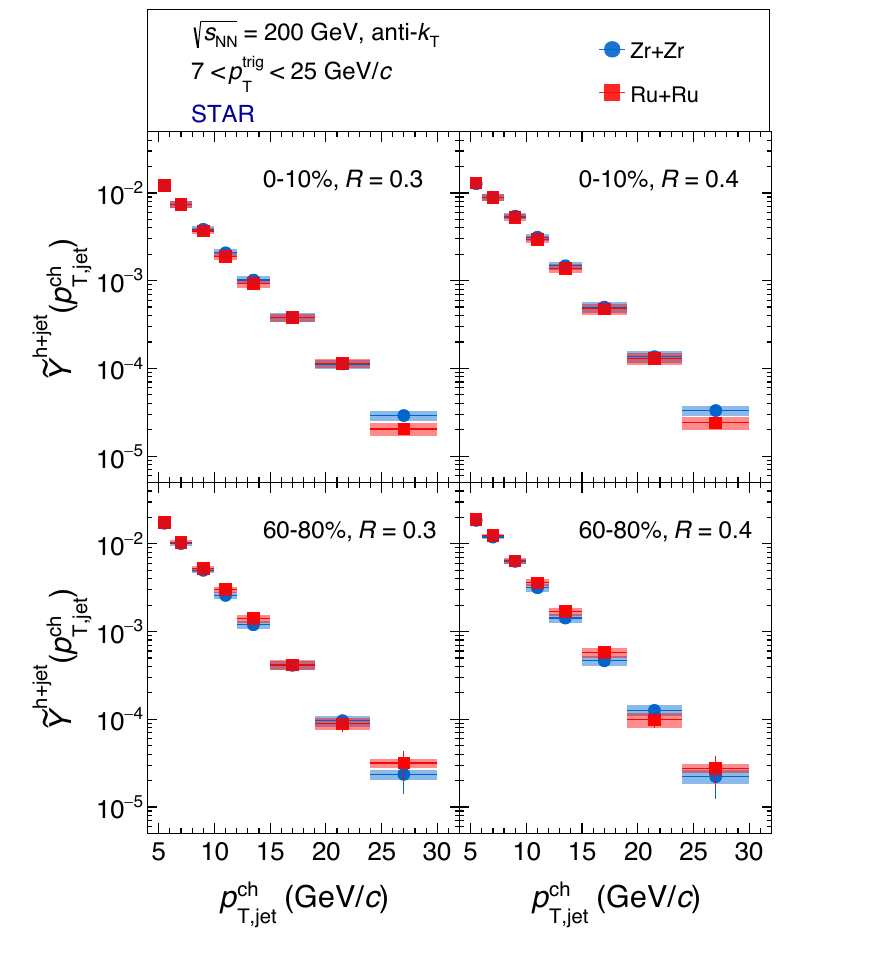}
\caption{Distributions of  \YtildehJet(\pTjetch)  for Zr+Zr and Ru+Ru collisions at \sqrtsNN\ = 200 GeV, for $R= 0.3$ (left) and 0.4 (right). Upper panels: central collisions; lower panels: peripheral collisions. Statistical errors and systematic uncertainties are indicated by error bars and shaded boxes, respectively.}
\label{fully correct spectra 2}
\end{figure}

Figure~\ref{fully correct spectra 2} shows distributions of \YtildehJet(\pTjetch) for central and peripheral Zr+Zr and Ru+Ru collisions at \sqrtsNN\ = 200 GeV, for $R= 0.3$ and 0.4.
\begin{figure}[hbtp]
	\centering
\includegraphics[width=0.49\textwidth]{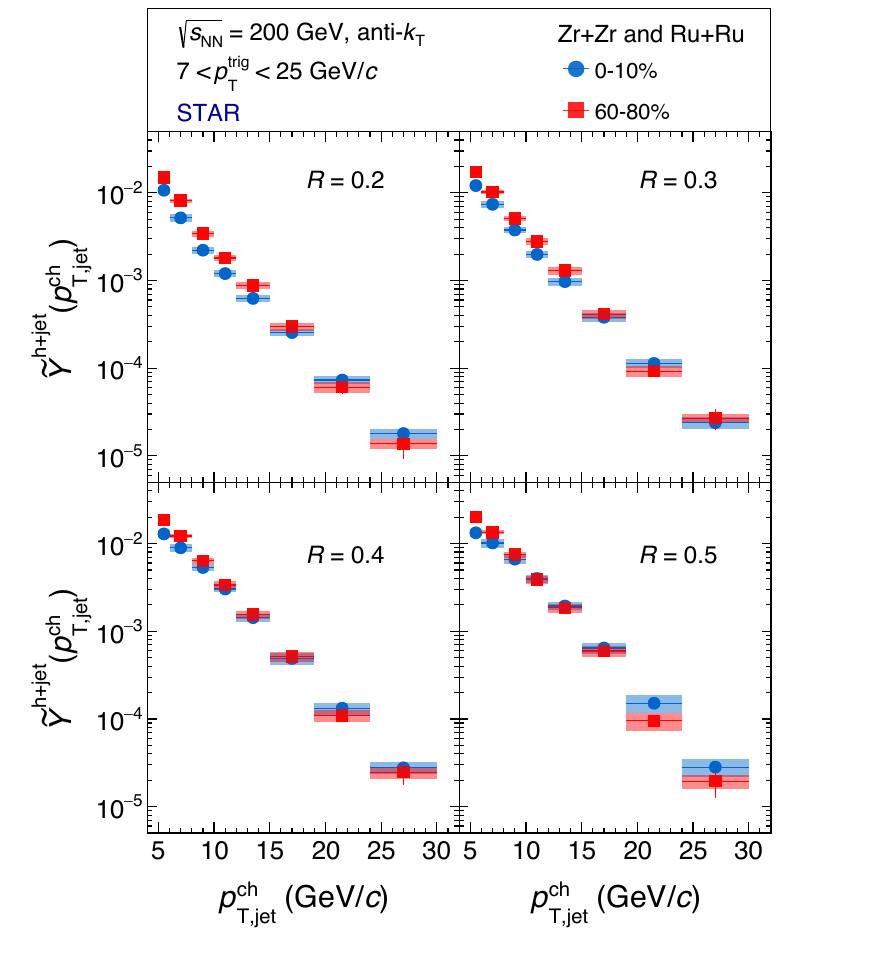}
\caption{Distributions of  \YtildehJet(\pTjetch) for the combined \RuRu\ and \ZrZr\ data, for central and peripheral collisions and $R=0.2$, 0.3, 0.4, and 0.5. Statistical errors and systematic uncertainties are indicated by error bars and shaded boxes, respectively.}
	\label{fully correct spectra combining}
\end{figure}
Fig.~\ref{fully correct spectra combining} shows \YtildehJet(\pTjetch) for the combined \RuRu\ and \ZrZr\ data.

\begin{figure}[htbp]
\centering
\includegraphics[width=0.4\textwidth]{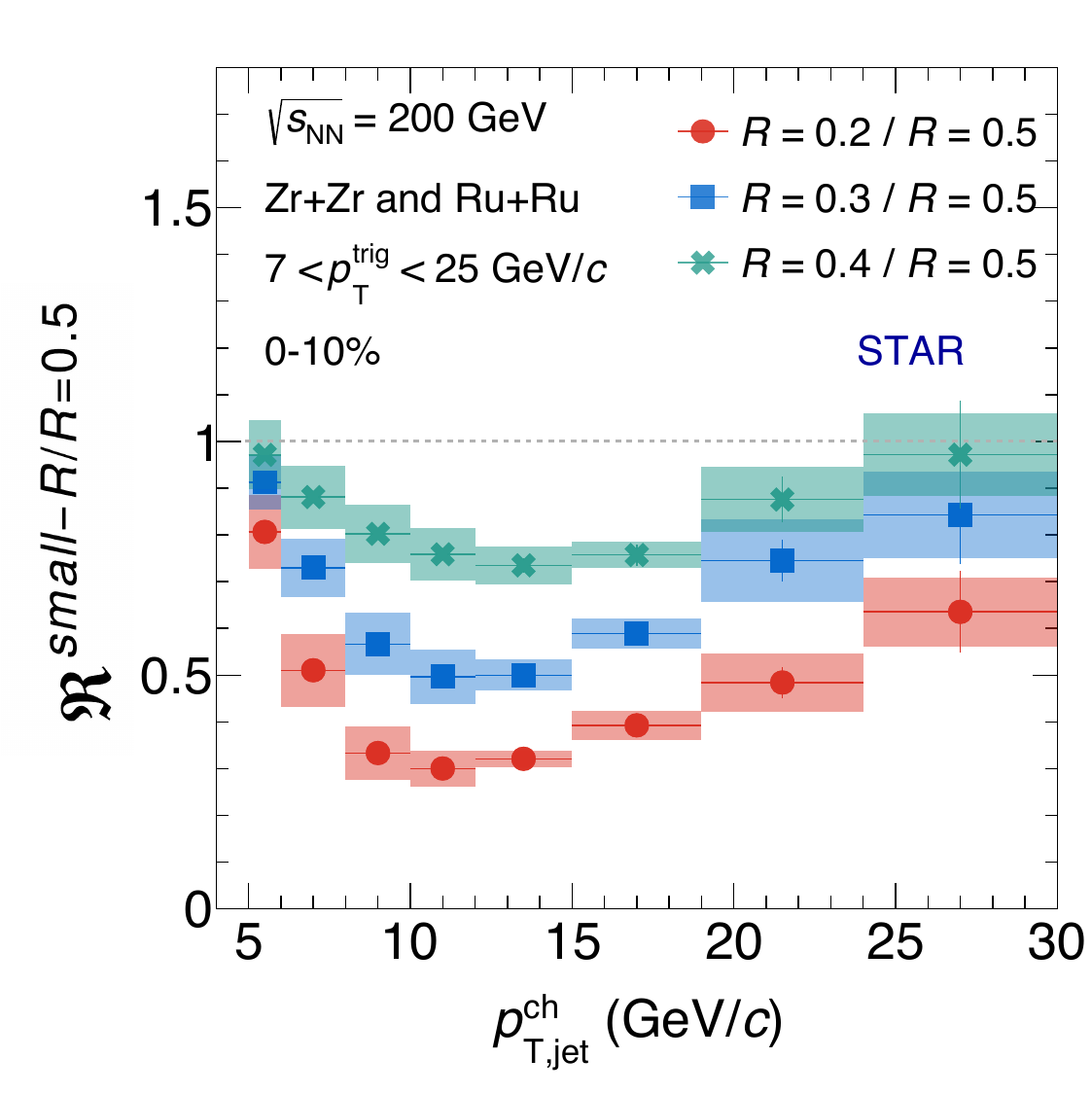}
\caption{Distributions of recoil jet yield ratios between small $R$ (0.2--0.4) and $R=0.5$ in 0--10\% central \RuRu\ and \ZrZr\ collisions at $\sqrtsNN=200$ GeV. }
\label{smallRfive}
\end{figure}

Figure \ref{smallRfive} shows ratios of recoil jet yields with small $R$ (0.2--0.4) to those of $R=0.5$ in 0--10\% central collisions. 


\begin{thebibliography}{10}
	\expandafter\ifx\csname url\endcsname\relax
	\def\url#1{\texttt{#1}}\fi
	\expandafter\ifx\csname urlprefix\endcsname\relax\def\urlprefix{URL }\fi
	\expandafter\ifx\csname href\endcsname\relax
	\def\href#1#2{#2} \def\path#1{#1}\fi
	
	\bibitem{Busza:2018rrf}
	W.~Busza, K.~Rajagopal, W.~van~der Schee, {Heavy Ion Collisions: The Big
		Picture, and the Big Questions}, Ann. Rev. Nucl. Part. Sci. 68 (2018)
	339--376.
	\newblock \href {http://arxiv.org/abs/1802.04801} {\path{arXiv:1802.04801}},
	\href {https://doi.org/10.1146/annurev-nucl-101917-020852}
	{\path{doi:10.1146/annurev-nucl-101917-020852}}.
	
	\bibitem{Elfner:2022iae}
	H.~Elfner, B.~M{\"u}ller, {The exploration of hot and dense nuclear matter:
		introduction to relativistic heavy-ion physics}, J. Phys. G 50~(10) (2023)
	103001.
	\newblock \href {http://arxiv.org/abs/2210.12056} {\path{arXiv:2210.12056}},
	\href {https://doi.org/10.1088/1361-6471/ace824}
	{\path{doi:10.1088/1361-6471/ace824}}.
	
	\bibitem{Harris:2023tti}
	J.~W. Harris, B.~M{\"u}ller, {''QGP Signatures'' Revisited}, Eur. Phys. J. C
	84~(3) (2024) 247.
	\newblock \href {http://arxiv.org/abs/2308.05743} {\path{arXiv:2308.05743}},
	\href {https://doi.org/10.1140/epjc/s10052-024-12533-y}
	{\path{doi:10.1140/epjc/s10052-024-12533-y}}.
	
	\bibitem{Schwarz:2003du}
	D.~J. Schwarz, {The first second of the universe}, Annalen Phys. 12 (2003)
	220--270.
	\newblock \href {http://arxiv.org/abs/astro-ph/0303574}
	{\path{arXiv:astro-ph/0303574}}, \href
	{https://doi.org/10.1002/andp.200310010} {\path{doi:10.1002/andp.200310010}}.
	
	\bibitem{Heinz:2013th}
	U.~Heinz, R.~Snellings, {Collective flow and viscosity in relativistic
		heavy-ion collisions}, Ann. Rev. Nucl. Part. Sci. 63 (2013) 123--151.
	\newblock \href {http://arxiv.org/abs/1301.2826} {\path{arXiv:1301.2826}},
	\href {https://doi.org/10.1146/annurev-nucl-102212-170540}
	{\path{doi:10.1146/annurev-nucl-102212-170540}}.
	
	\bibitem{2025-1834}
	J.~Chen, Z.~Chen, M.~Nie, H.~Qiu, S.~Shi, Z.~Tang, Q.~Xu, C.~Yang, S.~Yang,
	Z.~Ye, L.~Yi, W.~Zha, C.~Zhang, J.~Zhang, Y.~Zhang, X.~Zhu, {Selected
		highlights from STAR experiment}, Chin. Phys. Lett. 43~(3) (2026) 030102.
	\newblock \href {https://doi.org/10.1088/0256-307X/43/3/030102}
	{\path{doi:10.1088/0256-307X/43/3/030102}}.
	
	\bibitem{Sterman:1977wj}
	G.~F. Sterman, S.~Weinberg, {Jets from Quantum Chromodynamics}, Phys. Rev.
	Lett. 39 (1977) 1436.
	\newblock \href {https://doi.org/10.1103/PhysRevLett.39.1436}
	{\path{doi:10.1103/PhysRevLett.39.1436}}.
	
	\bibitem{Salam:2010nqg}
	G.~P. Salam, {Towards Jetography}, Eur. Phys. J. C 67 (2010) 637--686.
	\newblock \href {http://arxiv.org/abs/0906.1833} {\path{arXiv:0906.1833}},
	\href {https://doi.org/10.1140/epjc/s10052-010-1314-6}
	{\path{doi:10.1140/epjc/s10052-010-1314-6}}.
	
	\bibitem{Majumder:2010qh}
	A.~Majumder, M.~Van~Leeuwen, {The Theory and Phenomenology of Perturbative QCD
		Based Jet Quenching}, Prog. Part. Nucl. Phys. 66 (2011) 41--92.
	\newblock \href {http://arxiv.org/abs/1002.2206} {\path{arXiv:1002.2206}},
	\href {https://doi.org/10.1016/j.ppnp.2010.09.001}
	{\path{doi:10.1016/j.ppnp.2010.09.001}}.
	
	\bibitem{Wang:2025lct}
	X.-N. Wang, U.~A. Wiedemann, {QGP@50: More than Four Decades of Jet Quenching},
	2025.
	\newblock \href {http://arxiv.org/abs/2508.18794} {\path{arXiv:2508.18794}}.
	
	\bibitem{Cunqueiro:2021wls}
	L.~Cunqueiro, A.~M. Sickles, {Studying the QGP with Jets at the LHC and RHIC},
	Prog. Part. Nucl. Phys. 124 (2022) 103940.
	\newblock \href {http://arxiv.org/abs/2110.14490} {\path{arXiv:2110.14490}},
	\href {https://doi.org/10.1016/j.ppnp.2022.103940}
	{\path{doi:10.1016/j.ppnp.2022.103940}}.
	
	\bibitem{Apolinario:2022vzg}
	L.~Apolin{\'a}rio, Y.-J. Lee, M.~Winn, {Heavy quarks and jets as probes of the
		QGP}, Prog. Part. Nucl. Phys. 127 (2022) 103990.
	\newblock \href {http://arxiv.org/abs/2203.16352} {\path{arXiv:2203.16352}},
	\href {https://doi.org/10.1016/j.ppnp.2022.103990}
	{\path{doi:10.1016/j.ppnp.2022.103990}}.
	
	\bibitem{JET:2013cls}
	K.~M. Burke, et~al., {Extracting the jet transport coefficient from jet
		quenching in high-energy heavy-ion collisions}, Phys. Rev. C 90~(1) (2014)
	014909.
	\newblock \href {http://arxiv.org/abs/1312.5003} {\path{arXiv:1312.5003}},
	\href {https://doi.org/10.1103/PhysRevC.90.014909}
	{\path{doi:10.1103/PhysRevC.90.014909}}.
	
	\bibitem{JETSCAPE:2024cqe}
	R.~Ehlers, et~al., {Bayesian inference analysis of jet quenching using
		inclusive jet and hadron suppression measurements}, Phys. Rev. C 111~(5)
	(2025) 054913.
	\newblock \href {http://arxiv.org/abs/2408.08247} {\path{arXiv:2408.08247}},
	\href {https://doi.org/10.1103/PhysRevC.111.054913}
	{\path{doi:10.1103/PhysRevC.111.054913}}.
	
	\bibitem{Jing:2025bwi}
	P.~Jing, Y.~Dang, Y.~He, S.~Cao, L.~Yi, X.-N. Wang, {Emergence of thermal
		recoil jets in high-energy heavy-ion collisions} (12 2025).
	\newblock \href {http://arxiv.org/abs/2512.12715} {\path{arXiv:2512.12715}}.
	
	\bibitem{Landau:1965ksp}
	L.~D. Landau, I.~Pomeranchuk, {Electron-Cascade Processes at Ultra-High
		Energies}, Dokl. Akad. Nauk SSSR 92 (1965).
	\newblock \href {https://doi.org/10.1016/b978-0-08-010586-4.50081-x}
	{\path{doi:10.1016/b978-0-08-010586-4.50081-x}}.
	
	\bibitem{Migdal:1956tc}
	A.~B. Migdal, {Bremsstrahlung and Pair Production at High Energies in Condensed
		Media}, Phys. Rev. 103 (1956) 1811--1820.
	\newblock \href {https://doi.org/10.1103/PhysRev.103.1811}
	{\path{doi:10.1103/PhysRev.103.1811}}.
	
	\bibitem{He:2024xtk}
	Y.~He, M.~Zhang, M.~Nie, S.~Cao, L.~Yi, {Exploring system size dependence of
		jet modification in heavy-ion collisions}, Phys. Rev. C 110~(3) (2024)
	034902.
	\newblock \href {http://arxiv.org/abs/2404.18115} {\path{arXiv:2404.18115}},
	\href {https://doi.org/10.1103/PhysRevC.110.034902}
	{\path{doi:10.1103/PhysRevC.110.034902}}.
	
	\bibitem{ALICE:2012mj}
	B.~Abelev, et~al., {Transverse momentum distribution and nuclear modification
		factor of charged particles in $p$-Pb collisions at $\sqrt{s_{\mathrm {NN}}}$
		= 5.02 TeV}, Phys. Rev. Lett. 110~(8) (2013) 082302.
	\newblock \href {http://arxiv.org/abs/1210.4520} {\path{arXiv:1210.4520}},
	\href {https://doi.org/10.1103/PhysRevLett.110.082302}
	{\path{doi:10.1103/PhysRevLett.110.082302}}.
	
	\bibitem{ATLAS:2014cpa}
	G.~Aad, et~al., {Centrality and rapidity dependence of inclusive jet production
		in $\sqrt{s_{\mathrm {NN}}}$ = 5.02 TeV proton-lead collisions with the ATLAS
		detector}, Phys. Lett. B 748 (2015) 392--413.
	\newblock \href {http://arxiv.org/abs/1412.4092} {\path{arXiv:1412.4092}},
	\href {https://doi.org/10.1016/j.physletb.2015.07.023}
	{\path{doi:10.1016/j.physletb.2015.07.023}}.
	
	\bibitem{CMS:2016xef}
	V.~Khachatryan, et~al., {Charged-particle nuclear modification factors in PbPb
		and pPb collisions at $\sqrt{s_{\mathrm {NN}}}$ = 5.02 TeV}, JHEP 04 (2017)
	039.
	\newblock \href {http://arxiv.org/abs/1611.01664} {\path{arXiv:1611.01664}},
	\href {https://doi.org/10.1007/JHEP04(2017)039}
	{\path{doi:10.1007/JHEP04(2017)039}}.
	
	\bibitem{STAR:2024nwm}
	M.~Abdulhamid, et~al., {Correlations of event activity with hard and soft
		processes in p+Au collisions at $\sqrt{s_{\mathrm {NN}}}$ = 200 GeV at the
		RHIC STAR experiment}, Phys. Rev. C 110~(4) (2024) 044908.
	\newblock \href {http://arxiv.org/abs/2404.08784} {\path{arXiv:2404.08784}},
	\href {https://doi.org/10.1103/PhysRevC.110.044908}
	{\path{doi:10.1103/PhysRevC.110.044908}}.
	
	\bibitem{PHENIX:2023dxl}
	N.~J. Abdulameer, et~al., {Disentangling Centrality Bias and Final-State
		Effects in the Production of High-pT Neutral Pions Using Direct Photon in
		d+Au Collisions at $\sqrt{s_{\mathrm {NN}}}$ = 200 GeV}, Phys. Rev. Lett.
	134~(2) (2025) 022302.
	\newblock \href {http://arxiv.org/abs/2303.12899} {\path{arXiv:2303.12899}},
	\href {https://doi.org/10.1103/PhysRevLett.134.022302}
	{\path{doi:10.1103/PhysRevLett.134.022302}}.
	
	\bibitem{Perepelitsa:2024eik}
	D.~V. Perepelitsa, {Contribution to differential $\pi^{0}$ and $\gamma^{\rm
			dir}$ modification in small systems from color fluctuation effects}, Phys.
	Rev. C 110~(1) (2024) L011901.
	\newblock \href {http://arxiv.org/abs/2404.17660} {\path{arXiv:2404.17660}},
	\href {https://doi.org/10.1103/PhysRevC.110.L011901}
	{\path{doi:10.1103/PhysRevC.110.L011901}}.
	
	\bibitem{ALICE:2017svf}
	S.~Acharya, et~al., {Constraints on jet quenching in p-Pb collisions at
		$\sqrt{s_{\mathrm {NN}}}$ = 5.02 TeV measured by the event-activity
		dependence of semi-inclusive hadron-jet distributions}, Phys. Lett. B 783
	(2018) 95--113.
	\newblock \href {http://arxiv.org/abs/1712.05603} {\path{arXiv:1712.05603}},
	\href {https://doi.org/10.1016/j.physletb.2018.05.059}
	{\path{doi:10.1016/j.physletb.2018.05.059}}.
	
	\bibitem{ATLAS:2022iyq}
	G.~Aad, et~al., {Strong Constraints on Jet Quenching in Centrality-Dependent
		p+Pb Collisions at 5.02~TeV from ATLAS}, Phys. Rev. Lett. 131~(7) (2023)
	072301.
	\newblock \href {http://arxiv.org/abs/2206.01138} {\path{arXiv:2206.01138}},
	\href {https://doi.org/10.1103/PhysRevLett.131.072301}
	{\path{doi:10.1103/PhysRevLett.131.072301}}.
	
	\bibitem{CMS:2025jbv}
	V.~Chekhovsky, et~al., {Search for jet quenching with dijets from
		high-multiplicity pPb collisions at $\sqrt{s_{\mathrm {NN}}}$ = 8.16 TeV},
	JHEP 07 (2025) 118.
	\newblock \href {http://arxiv.org/abs/2504.08507} {\path{arXiv:2504.08507}},
	\href {https://doi.org/10.1007/JHEP07(2025)118}
	{\path{doi:10.1007/JHEP07(2025)118}}.
	
	\bibitem{CMS:2025bta}
	A.~Hayrapetyan, et~al., {Discovery of suppressed charged-particle production in
		ultrarelativistic oxygen-oxygen collisions} (2025).
	\newblock \href {http://arxiv.org/abs/2510.09864} {\path{arXiv:2510.09864}}.
	
	\bibitem{STAR:2026nfy}
	{Measurement of jet quenching in O+O collisions at $\sqrt{s_\mathrm{NN}}=200$
		GeV by the STAR experiment at RHIC} (2026).
	\newblock \href {http://arxiv.org/abs/2604.13935} {\path{arXiv:2604.13935}}.
	
	\bibitem{STAR:2021mii}
	M.~Abdallah, et~al., {Search for the chiral magnetic effect with isobar
		collisions at $\sqrt{s_{\mathrm {NN}}}$ = 200 GeV by the STAR Collaboration
		at the BNL Relativistic Heavy Ion Collider}, Phys. Rev. C 105~(1) (2022)
	014901.
	\newblock \href {http://arxiv.org/abs/2109.00131} {\path{arXiv:2109.00131}},
	\href {https://doi.org/10.1103/PhysRevC.105.014901}
	{\path{doi:10.1103/PhysRevC.105.014901}}.
	
	\bibitem{ALICE:2015mdb}
	J.~Adam, et~al., {Measurement of jet quenching with semi-inclusive hadron-jet
		distributions in central Pb-Pb collisions at $\sqrt{s_{\mathrm {NN}}}$ = 2.76
		TeV}, JHEP 09 (2015) 170.
	\newblock \href {http://arxiv.org/abs/1506.03984} {\path{arXiv:1506.03984}},
	\href {https://doi.org/10.1007/JHEP09(2015)170}
	{\path{doi:10.1007/JHEP09(2015)170}}.
	
	\bibitem{STAR:2017hhs}
	L.~Adamczyk, et~al., {Measurements of jet quenching with semi-inclusive
		hadron+jet distributions in Au+Au collisions at $\sqrt{s_{\mathrm {NN}}}$ =
		200 GeV}, Phys. Rev. C 96~(2) (2017) 024905.
	\newblock \href {http://arxiv.org/abs/1702.01108} {\path{arXiv:1702.01108}},
	\href {https://doi.org/10.1103/PhysRevC.96.024905}
	{\path{doi:10.1103/PhysRevC.96.024905}}.
	
	\bibitem{ALICE:2019qyj}
	S.~Acharya, et~al., {Measurements of inclusive jet spectra in pp and central
		Pb-Pb collisions at $\sqrt{s_{\rm{NN}}}$ = 5.02 TeV}, Phys. Rev. C 101~(3)
	(2020) 034911.
	\newblock \href {http://arxiv.org/abs/1909.09718} {\path{arXiv:1909.09718}},
	\href {https://doi.org/10.1103/PhysRevC.101.034911}
	{\path{doi:10.1103/PhysRevC.101.034911}}.
	
	\bibitem{ALICE:2023jye}
	S.~Acharya, et~al., {Measurements of jet quenching using semi-inclusive
		hadron+jet distributions in pp and central Pb-Pb collisions at
		$\sqrt{s_{\mathrm {NN}}}$ = 5.02 TeV}, Phys. Rev. C 110~(1) (2024) 014906.
	\newblock \href {http://arxiv.org/abs/2308.16128} {\path{arXiv:2308.16128}},
	\href {https://doi.org/10.1103/PhysRevC.110.014906}
	{\path{doi:10.1103/PhysRevC.110.014906}}.
	
	\bibitem{ALICE:2023qve}
	S.~Acharya, et~al., {Observation of Medium-Induced Yield Enhancement and
		Acoplanarity Broadening of Low-$p_T$ Jets from Measurements in pp and Central
		Pb-Pb Collisions at $\sqrt{s_{\mathrm {NN}}}$ = 5.02 TeV}, Phys. Rev. Lett.
	133~(2) (2024) 022301.
	\newblock \href {http://arxiv.org/abs/2308.16131} {\path{arXiv:2308.16131}},
	\href {https://doi.org/10.1103/PhysRevLett.133.022301}
	{\path{doi:10.1103/PhysRevLett.133.022301}}.
	
	\bibitem{STAR:2023ksv}
	B.~E. Aboona, et~al., {Semi-inclusive direct photon+jet and $\pi^{0}$+jet
		correlations measured in p+p and central Au+Au collisions at
		$\sqrt{s_{\mathrm {NN}}}$ = 200 GeV}, Phys. Rev. C 111~(6) (2025) 064907.
	\newblock \href {http://arxiv.org/abs/2309.00145} {\path{arXiv:2309.00145}},
	\href {https://doi.org/10.1103/8b8y-98yh} {\path{doi:10.1103/8b8y-98yh}}.
	
	\bibitem{STAR:2023pal}
	B.~E. Aboona, et~al., {Measurement of In-Medium Jet Modification Using Direct
		Photon+Jet and $\pi^{0}$+Jet Correlations in p+p and Central Au+Au Collisions
		at $\sqrt{s_{\mathrm {NN}}}$ = 200 GeV}, Phys. Rev. Lett. 134~(23) (2025)
	232301.
	\newblock \href {http://arxiv.org/abs/2309.00156} {\path{arXiv:2309.00156}},
	\href {https://doi.org/10.1103/PhysRevLett.134.232301}
	{\path{doi:10.1103/PhysRevLett.134.232301}}.
	
	\bibitem{STAR:2025yhg}
	B.~E. Aboona, et~al., {Measurement of medium-induced acoplanarity in central
		Au-Au and pp collisions at $\sqrt{s_{\mathrm {NN}}}$ = 200 GeV using
		direct-photon+jet and $\pi^{0}$+jet correlations}, Phys. Rev. C 113~(1)
	(2026) 014902.
	\newblock \href {http://arxiv.org/abs/2505.05789} {\path{arXiv:2505.05789}},
	\href {https://doi.org/10.1103/k29c-d5ry} {\path{doi:10.1103/k29c-d5ry}}.
	
	\bibitem{STAR:2002eio}
	K.~H. Ackermann, et~al., {STAR detector overview}, Nucl. Instrum. Meth. A 499
	(2003) 624--632.
	\newblock \href {https://doi.org/10.1016/S0168-9002(02)01960-5}
	{\path{doi:10.1016/S0168-9002(02)01960-5}}.
	
	\bibitem{Anderson:2003ur}
	M.~Anderson, et~al., {The Star time projection chamber: A Unique tool for
		studying high multiplicity events at RHIC}, Nucl. Instrum. Meth. A 499 (2003)
	659--678.
	\newblock \href {http://arxiv.org/abs/nucl-ex/0301015}
	{\path{arXiv:nucl-ex/0301015}}, \href
	{https://doi.org/10.1016/S0168-9002(02)01964-2}
	{\path{doi:10.1016/S0168-9002(02)01964-2}}.
	
	\bibitem{Llope:2005yw}
	W.~J. Llope, {The large-area time-of-flight upgrade for STAR}, Nucl. Instrum.
	Meth. B 241 (2005) 306--310.
	\newblock \href {https://doi.org/10.1016/j.nimb.2005.07.089}
	{\path{doi:10.1016/j.nimb.2005.07.089}}.
	
	\bibitem{Llope:2014nva}
	W.~J. Llope, et~al., {The STAR Vertex Position Detector}, Nucl. Instrum. Meth.
	A 759 (2014) 23--28.
	\newblock \href {http://arxiv.org/abs/1403.6855} {\path{arXiv:1403.6855}},
	\href {https://doi.org/10.1016/j.nima.2014.04.080}
	{\path{doi:10.1016/j.nima.2014.04.080}}.
	
	\bibitem{Adler:2003sp}
	C.~Adler, A.~Denisov, E.~Garcia, M.~Murray, H.~Strobele, S.~White, {The RHIC
		zero-degree calorimeters}, Nucl. Instrum. Meth. A 499 (2003) 433--436.
	\newblock \href {https://doi.org/10.1016/j.nima.2003.08.112}
	{\path{doi:10.1016/j.nima.2003.08.112}}.
	
	\bibitem{Miller:2007ri}
	M.~L. Miller, K.~Reygers, S.~J. Sanders, P.~Steinberg, {Glauber modeling in
		high energy nuclear collisions}, Ann. Rev. Nucl. Part. Sci. 57 (2007)
	205--243.
	\newblock \href {http://arxiv.org/abs/nucl-ex/0701025}
	{\path{arXiv:nucl-ex/0701025}}, \href
	{https://doi.org/10.1146/annurev.nucl.57.090506.123020}
	{\path{doi:10.1146/annurev.nucl.57.090506.123020}}.
	
	\bibitem{Kalman:1960mft}
	R.~E. Kalman, {A New Approach to Linear Filtering and Prediction Problems}, J.
	Fluids Eng. 82~(1) (1960) 35--45.
	\newblock \href {https://doi.org/10.1115/1.3662552}
	{\path{doi:10.1115/1.3662552}}.
	
	\bibitem{Fruhwirth:1987fm}
	R.~Fruhwirth, {Application of Kalman filtering to track and vertex fitting},
	Nucl. Instrum. Meth. A 262 (1987) 444--450.
	\newblock \href {https://doi.org/10.1016/0168-9002(87)90887-4}
	{\path{doi:10.1016/0168-9002(87)90887-4}}.
	
	\bibitem{Catani:1993hr}
	S.~Catani, Y.~L. Dokshitzer, M.~H. Seymour, B.~R. Webber, {Longitudinally
		invariant $K_t$ clustering algorithms for hadron hadron collisions}, Nucl.
	Phys. B 406 (1993) 187--224.
	\newblock \href {https://doi.org/10.1016/0550-3213(93)90166-M}
	{\path{doi:10.1016/0550-3213(93)90166-M}}.
	
	\bibitem{Cacciari:2005hq}
	M.~Cacciari, G.~P. Salam, {Dispelling the $N^{3}$ myth for the $k_t$
		jet-finder}, Phys. Lett. B 641 (2006) 57--61.
	\newblock \href {http://arxiv.org/abs/hep-ph/0512210}
	{\path{arXiv:hep-ph/0512210}}, \href
	{https://doi.org/10.1016/j.physletb.2006.08.037}
	{\path{doi:10.1016/j.physletb.2006.08.037}}.
	
	\bibitem{Cacciari:2011ma}
	M.~Cacciari, G.~P. Salam, G.~Soyez, {FastJet User Manual}, Eur. Phys. J. C 72
	(2012) 1896.
	\newblock \href {http://arxiv.org/abs/1111.6097} {\path{arXiv:1111.6097}},
	\href {https://doi.org/10.1140/epjc/s10052-012-1896-2}
	{\path{doi:10.1140/epjc/s10052-012-1896-2}}.
	
	\bibitem{Cacciari:2007fd}
	M.~Cacciari, G.~P. Salam, {Pileup subtraction using jet areas}, Phys. Lett. B
	659 (2008) 119--126.
	\newblock \href {http://arxiv.org/abs/0707.1378} {\path{arXiv:0707.1378}},
	\href {https://doi.org/10.1016/j.physletb.2007.09.077}
	{\path{doi:10.1016/j.physletb.2007.09.077}}.
	
	\bibitem{Cacciari:2008gp}
	M.~Cacciari, G.~P. Salam, G.~Soyez, {The anti-$k_t$ jet clustering algorithm},
	JHEP 04 (2008) 063.
	\newblock \href {http://arxiv.org/abs/0802.1189} {\path{arXiv:0802.1189}},
	\href {https://doi.org/10.1088/1126-6708/2008/04/063}
	{\path{doi:10.1088/1126-6708/2008/04/063}}.
	
	\bibitem{Cacciari:2010te}
	M.~Cacciari, J.~Rojo, G.~P. Salam, G.~Soyez, {Jet Reconstruction in Heavy Ion
		Collisions}, Eur. Phys. J. C 71 (2011) 1539.
	\newblock \href {http://arxiv.org/abs/1010.1759} {\path{arXiv:1010.1759}},
	\href {https://doi.org/10.1140/epjc/s10052-011-1539-z}
	{\path{doi:10.1140/epjc/s10052-011-1539-z}}.
	
	\bibitem{Brenner:2019lmf}
	L.~Brenner, R.~Balasubramanian, C.~Burgard, W.~Verkerke, G.~Cowan,
	P.~Verschuuren, V.~Croft, {Comparison of unfolding methods using
		RooFitUnfold}, Int. J. Mod. Phys. A 35~(24) (2020) 2050145.
	\newblock \href {http://arxiv.org/abs/1910.14654} {\path{arXiv:1910.14654}},
	\href {https://doi.org/10.1142/S0217751X20501456}
	{\path{doi:10.1142/S0217751X20501456}}.
	
	\bibitem{Adkins:2015ccl}
	J.~K. Adkins, {Studying Transverse Momentum Dependent Distributions in
		Polarized Proton Collisions via Azimuthal Single Spin Asymmetries of Charged
		Pions in Jets}, Ph.D. thesis, Kentucky U. (2015).
	\newblock \href {http://arxiv.org/abs/1907.11233} {\path{arXiv:1907.11233}}.
	
	\bibitem{DAgostini:1994fjx}
	G.~D'Agostini, {A Multidimensional unfolding method based on Bayes' theorem},
	Nucl. Instrum. Meth. A 362 (1995) 487--498.
	\newblock \href {https://doi.org/10.1016/0168-9002(95)00274-X}
	{\path{doi:10.1016/0168-9002(95)00274-X}}.
	
	\bibitem{Roounf}
	{RooUnfold}, {Root Unfolding Framework},
	\url{https://gitlab.cern.ch/RooUnfold/RooUnfold}.
	
	\bibitem{Dang:2026ezw}
	Y.~Dang, W.-J. Xing, S.~Cao, G.-Y. Qin, {Improved linear Boltzmann transport
		model for hadron and jet suppression in ultra-relativistic heavy-ion
		collisions} (2 2026).
	\newblock \href {http://arxiv.org/abs/2602.10395} {\path{arXiv:2602.10395}}.
	
	\bibitem{privatecommLBT}
	P.~Jing, S.~Cao, Private communication (2026).
	
	\bibitem{Baier:2002tc}
	R.~Baier, {Jet quenching}, Nucl. Phys. A 715 (2003) 209--218.
	\newblock \href {http://arxiv.org/abs/hep-ph/0209038}
	{\path{arXiv:hep-ph/0209038}}, \href
	{https://doi.org/10.1016/S0375-9474(02)01429-X}
	{\path{doi:10.1016/S0375-9474(02)01429-X}}.
	
	\bibitem{Dainese:2004te}
	A.~Dainese, C.~Loizides, G.~Paic, {Leading-particle suppression in high energy
		nucleus-nucleus collisions}, Eur. Phys. J. C 38 (2005) 461--474.
	\newblock \href {http://arxiv.org/abs/hep-ph/0406201}
	{\path{arXiv:hep-ph/0406201}}, \href
	{https://doi.org/10.1140/epjc/s2004-02077-x}
	{\path{doi:10.1140/epjc/s2004-02077-x}}.
	
	\bibitem{Renk:2006nd}
	T.~Renk, {High ${p}_{T}$ hadrons as probes of the central region of Au-Au
		collisions at $\sqrt{s_{\mathrm {NN}}}$ = 200 GeV}, Phys. Rev. C 74 (2006)
	024903.
	\newblock \href {http://arxiv.org/abs/hep-ph/0602045}
	{\path{arXiv:hep-ph/0602045}}, \href
	{https://doi.org/10.1103/PhysRevC.74.024903}
	{\path{doi:10.1103/PhysRevC.74.024903}}.
	
	\bibitem{Zhang:2007ja}
	H.~Zhang, J.~F. Owens, E.~Wang, X.-N. Wang, {Dihadron tomography of high-energy
		nuclear collisions in NLO pQCD}, Phys. Rev. Lett. 98 (2007) 212301.
	\newblock \href {http://arxiv.org/abs/nucl-th/0701045}
	{\path{arXiv:nucl-th/0701045}}, \href
	{https://doi.org/10.1103/PhysRevLett.98.212301}
	{\path{doi:10.1103/PhysRevLett.98.212301}}.
	
	\bibitem{Cao:2016gvr}
	S.~Cao, T.~Luo, G.-Y. Qin, X.-N. Wang, {Linearized Boltzmann transport model
		for jet propagation in the quark-gluon plasma: Heavy quark evolution}, Phys.
	Rev. C 94~(1) (2016) 014909.
	\newblock \href {http://arxiv.org/abs/1605.06447} {\path{arXiv:1605.06447}},
	\href {https://doi.org/10.1103/PhysRevC.94.014909}
	{\path{doi:10.1103/PhysRevC.94.014909}}.
	
	\bibitem{Luo:2023nsi}
	T.~Luo, Y.~He, S.~Cao, X.-N. Wang, {Linear Boltzmann transport for jet
		propagation in the quark-gluon plasma: Inelastic processes and jet
		modification}, Phys. Rev. C 109~(3) (2024) 034919.
	\newblock \href {http://arxiv.org/abs/2306.13742} {\path{arXiv:2306.13742}},
	\href {https://doi.org/10.1103/PhysRevC.109.034919}
	{\path{doi:10.1103/PhysRevC.109.034919}}.
	
	\bibitem{He:2024rcv}
	Y.~He, M.~Nie, S.~Cao, R.~Ma, L.~Yi, H.~Caines, {Deciphering yield modification
		of hadron-triggered semi-inclusive recoil jets in heavy-ion collisions},
	Phys. Lett. B 854 (2024) 138739.
	\newblock \href {http://arxiv.org/abs/2401.05238} {\path{arXiv:2401.05238}},
	\href {https://doi.org/10.1016/j.physletb.2024.138739}
	{\path{doi:10.1016/j.physletb.2024.138739}}.
	
	\bibitem{Zhang:2022ctd}
	M.~Zhang, Y.~He, S.~Cao, L.~Yi, {Effects of the formation time of parton shower
		on jet quenching in heavy-ion collisions}, Chin. Phys. C 47~(2) (2023)
	024106.
	\newblock \href {http://arxiv.org/abs/2208.13331} {\path{arXiv:2208.13331}},
	\href {https://doi.org/10.1088/1674-1137/aca4c1}
	{\path{doi:10.1088/1674-1137/aca4c1}}.
	
	\bibitem{STAR:2016jdz}
	L.~Adamczyk, et~al., {Jet-like Correlations with Direct-Photon and Neutral-Pion
		Triggers at $\sqrt{s_{_{NN}}} = 200$ GeV}, Phys. Lett. B 760 (2016) 689--696.
	\newblock \href {http://arxiv.org/abs/1604.01117} {\path{arXiv:1604.01117}},
	\href {https://doi.org/10.1016/j.physletb.2016.07.046}
	{\path{doi:10.1016/j.physletb.2016.07.046}}.
	
\end{thebibliography}

\end{document}